\documentclass[conference,a4paper]{IEEEtran}
\usepackage[utf8]{inputenc} 
\usepackage[T1]{fontenc}
\usepackage{url}
\usepackage{ifthen}
\usepackage{cite}
\usepackage[cmex10]{amsmath}

\usepackage{graphicx}
\usepackage{amsfonts}
\usepackage{enumitem}
\usepackage{amssymb}
\usepackage{mathrsfs}
\usepackage{bm}
\usepackage{mdwmath}
\usepackage{subfigure}
\usepackage{mdwtab}
\usepackage{dsfont}
\usepackage[ruled,vlined,linesnumbered]{algorithm2e} 

\DeclareMathOperator*{\argmin}{arg\,min}

\newtheorem{lemma}{Lemma}

\newtheorem{defi}{Definition}

\newenvironment{iarray}{\begin{IEEEeqnarray}{rCl}}{\end{IEEEeqnarray}\ignorespacesafterend}

\begin{document}
    \title{Status from a Random Field: How Densely Should One Update?}
    \author{\IEEEauthorblockN{Zhiyuan Jiang}
    	\IEEEauthorblockA{Shanghai University, Shanghai 200444, China\\
    		Email: jiangzhiyuan@shu.edu.cn}
    	\and
    	\IEEEauthorblockN{Sheng Zhou}
    	\IEEEauthorblockA{Tsinghua University, Beijing 100084, China\\
    		Email: sheng.zhou@tsinghua.edu.cn}
    	}
%    	\IEEEauthorblockN{Zhiyuan Jiang$^1$, Sheng Zhou$^2$}
%    \IEEEauthorblockA{$^1$ jiangzhiyuan@shu.edu.cn, Shanghai University, Shanghai 200444, China\\
%        $^2$ sheng.zhou@tsinghua.edu.cn, Tsinghua University, Beijing 100084, China}}

    \maketitle
    
    \begin{abstract}
    In many applications, status information of a general spatial process, in contrast to a point information source, is of interest. In this paper, we consider a system where status information is drawn from a random field and transmitted to a fusion center through a wireless multiaccess channel. The optimal density of spatial sampling points to minimize the remote status estimation error is investigated. Assuming a one-dimensional Gauss Markov random field and an exponential correlation function, closed-form expressions of remote estimation error are obtained for First-Come First-Served (FCFS) and Last-Come First-Served (LCFS) service disciplines. The optimal spatial sampling density for the LCFS case is given explicitly. Simulation results are presented which agree with our analysis.
    \end{abstract}
    
    \section{Introduction}
    \label{sec_intro}    
    Accurate remote estimations of information about physical world have long been considered to have high practical value. Typical applications include e.g., temperature and air quality monitoring in the cities \cite{villa16}, random sensor deployment from the sky \cite{oll07} for unreachable and hazardous open environments, and channel state information database for efficient wireless communications \cite{deng18}. For many such applications, information estimations with high accuracy need to be maintained at a fusion center, usually achieved by information update packets transmitted from distributed sources to the fusion center wirelessly. The design and analysis of the information update system are thus of significant importance.
    
    Considering time-varying status information, it is natural to optimize the system towards keeping the freshest possible information estimations at the fusion center, based on the intuition that stale information is less accurate. The concept of \emph{Age of Information} (AoI) \cite{kaul12,costa16,najm17,sun17} is tailored to quantize the information staleness, which is defined as the elapsed time since the generation of the last successful update from an information source. For a stationary Markov source, whose future information is independent of the past given the present and its information statistics does not change when shifted in time, AoI can completely characterize the information estimation accuracy as shown by Sun \emph{et al.} \cite{yin18}, in the sense that the mutual information between estimations and original information can be described by a function of AoI only. Considerable amount of efforts have been put into analyzing and optimizing AoI in various scenarios, among which, Kaul \emph{et al.} \cite{kaul12} utilized a queuing analysis to show that there exists an optimal time-domain sampling frequency for AoI minimization which balances the tradeoff between producing too many updates to jam the queue, and sampling too sparsely hence losing track of the information variation. 
    
    On the other hand, in many cases, not only timely updates are required, which is equivalent to appropriately frequent time-domain sampling, but optimized spatial-domain sampling is also needed. On account of status from---instead of a single source of interest or several i.i.d. sources---a continuous time-spatial process with certain spatial structure, spatial-domain sampling also faces the problem of sampling density considering network queuing and scheduling delay. For example, consider the temperature statuses from an area transmitted through a common wireless channel to the fusion center; the statuses are obviously spatially correlated and hence strong spatial structure exists. Therefore, similar with the time-domain sampling tradeoff, the spatial sampling density should be optimized---a high density leads to high scheduling delay whereas a low density is insufficient to describe the area.
    \begin{figure}[!t]
    	\centering    
    	{\includegraphics[width=0.5\textwidth]{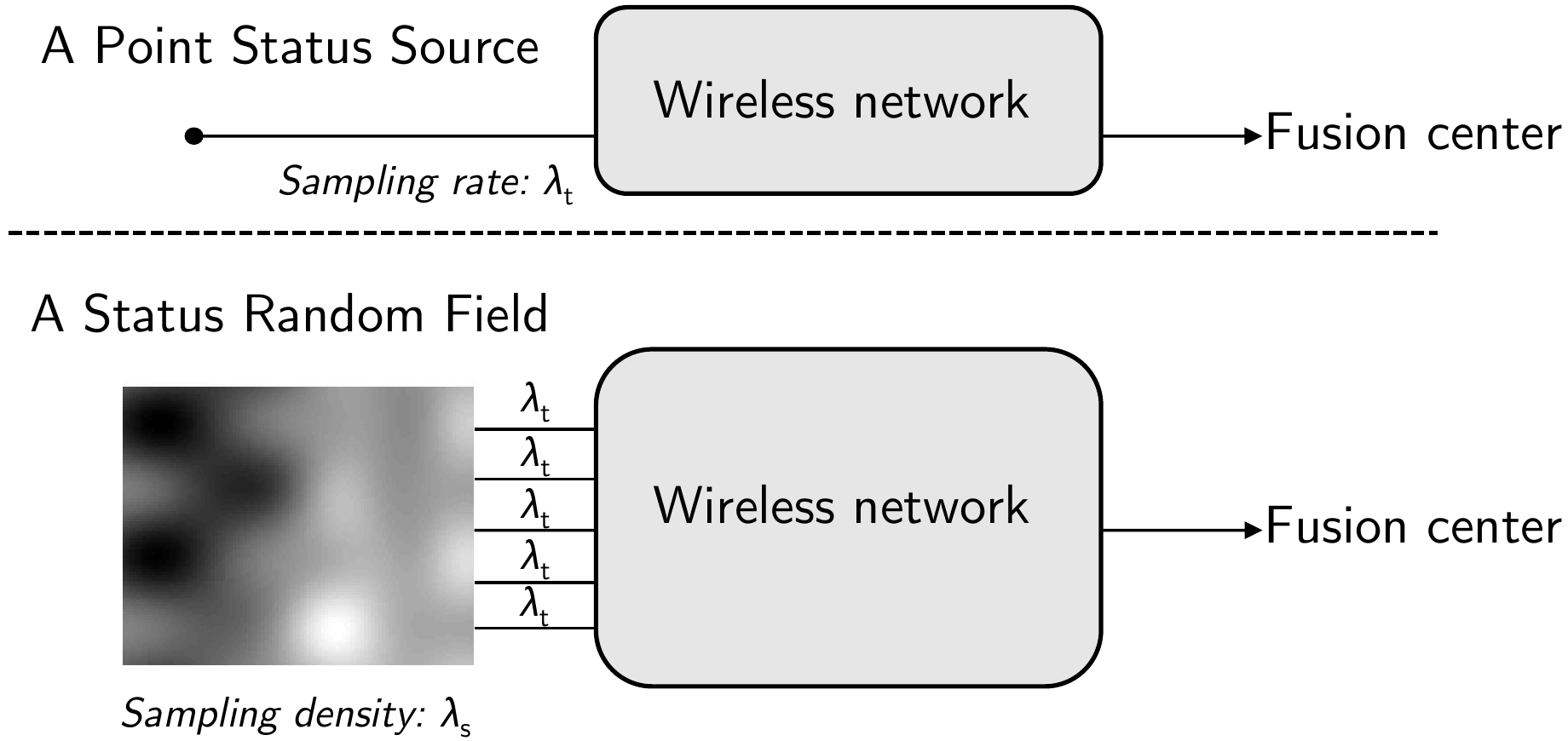}}
    	\caption{Differences between status update procedures where statuses are drawn from a point source and a random field, respectively.}
    	\label{fig_arch}
    \end{figure}

    From a more general perspective, this paper aims to investigate the problem of remote estimations through a channel with limited capacity, about information with general time-spatial structures. The differences of this work from previous ones are illustrated in Fig. \ref{fig_arch}. We focus on status information here, i.e., information with Markov time-domain structure, and a specific spatial correlation model. The main contributions include:
    
    1) The information spatial structure is modeled as a Gaussian Markov Random Field (GMRF). The spatial sampling points are deployed based on a homogeneous Poisson point process with density $\lambda_\mathsf{s}$. The information estimation error of some point in the GMRF is formulated as a function of its distance to the nearest sampling point (Markov property) and the corresponding AoI.
    
    2) The time- and spatial-average remote estimation error in a one-dimensional GMRF with the First-Come First-Served (FCFS) discipline and an exponential correlation function is derived in a closed-form. The optimal time- and spatial-sampling rates can be found by a two-dimensional search. 
    
    3) For Last-Come First-Served (LCFS) service discipline, closed-form remote estimation error expressions and the optimal time- and spatial-domain sampling rates are obtained.
      
    Among the recent progress, works that consider the scheduling of multiple points for AoI optimization are the most relevant, e.g., in \cite{kadota18,hsu18,jiang18_iot,jiang18_itc,rajat19,jiang19}. Hsu \emph{et al.} considered the scheduling problem in wireless broadcast channels and proved that a AoI threshold-based approach is optimal. The Whittle's index was leveraged by Kadota \emph{et al.} \cite{kadota18} based on a restless multi-armed-bandit formulation assuming active sources. In our previous works \cite{jiang18_itc,jiang18_iot,jiang19}, status update through a wireless multiaccess network was studied wherein the Whittle's index based near-optimal scheme and closed-form analysis were obtained. However, these works assumed i.i.d. status sources and thus the results are not generalizable to status from a correlated random field. Hribar \emph{et al.} \cite{hri18} considered correlated sources wherein an extremely simplified scenario consisting of two sources were studied. 
    
    \section{System Model and Problem Formulations}
    \label{sec_sm}
    Consider a random field on $\mathcal{A} \subset \mathbb{R}^d$, with the Lebesgue measure (i.e., volume) of $\mathcal{A}$ denoted by $|\mathcal{A}|$, i.e.,
    \begin{equation}
    S(\boldsymbol{x},t): \boldsymbol{x} \in \mathcal{A}, t\in \mathbb{R}^+,
    \end{equation}
    where we consider $S(\boldsymbol{x},t)$ is a real-valued stochastic process defined over space and time. At a remote fusion center, denote the information estimation of $S(\boldsymbol{x},t)$ as $\tilde{S}(\boldsymbol{x},t)$, and it is desirable to recover $S(\boldsymbol{x},t)$ for any $\boldsymbol{x},t$, i.e., at any spatial location and time, with high accuracy. Towards this end, status updates from the random field are transmitted to the fusion center through a wireless multiaccess channel. The status packets go though the channel, experiencing stochastic queuing and scheduling delay. 
    
    \textbf{Uniformly random spatial sampling}: We assume that the spatial sampling points are deployed in the random field in a uniformly random manner with a density of $\lambda_\mathsf{s}$ (a total of $M=\lambda_\mathsf{s} |\mathcal{A}|$ points). Neglecting edge effects, or equivalently considering a large enough area of $\mathcal{A}$ ($|\mathcal{A}| \to \infty$), the sampling points obey homogeneous Poisson point process \cite{baccelli10}, and the set of spatial sampling points are denoted by $\boldsymbol{X} \triangleq \{\boldsymbol{x}_1,\boldsymbol{x}_1,\cdots,\boldsymbol{x}_M\}$; that is, The Cumulative Distribution Function (CDF) of the distance (denoted by $D$) between some point in $\mathcal{A}$ and its nearest sampling point is
    \begin{equation}
    \Pr\{D \le r\} = \lim\limits_{|\mathcal{A}| \to \infty} 1-\left(\frac{|B(r)|}{|\mathcal{A}|}\right)^{M} = 1-\exp{(-\lambda_\mathsf{s} |B(r)|)},
    \end{equation}
    where $B(r)$ denotes a $d$-dimensional ball with a radius of $r$. 
    
    \textbf{Queuing and scheduling models}: In this paper, for mathematical tractability, we assume that the time for transmitting one sample over the wireless channel is i.i.d. exponentially distributed with mean $1/\mu$. Define the normalized (by spatial domain) service rate as $\bar{\mu} \triangleq \lim\limits_{|\mathcal{A}| \to \infty}\frac{\mu}{|\mathcal{A}|}$. Note that by definition $\mu \to \infty$; in practice, a large $\mu$ is sufficient. The update packets are queued at each spatial sampling point with an identical packet arrival rate of $\lambda_\mathsf{t}$; in this paper, we assume the status variation in time is homogeneous in the spatial domain, and thus the packet arrivals at different spatial points are statistically identical. Upon each transmission, only the status packet from one sampling point $\boldsymbol{x}_m \in \boldsymbol{X}$ is updated. Here, we consider a specific scheduling among status updates of spatial sampling points:
    \begin{itemize}
    	\item
    	\emph{Uniformly random scheduling}: Albeit suboptimal \cite{jiang18_iot}, uniformly random scheduling represents a scenario wherein nodes undergo a random access process with equal priority and then transmit, and thus of practical value.  
    \end{itemize}
    The average service time for each sampling point is $\frac{1}{\mu_0} \triangleq \frac{M}{\mu}=\frac{\lambda_\mathsf{s}}{\bar{\mu}}$.
    
    \textbf{Time and spatial structures of information}: A normalized isotropic (correlation only depends on distance, not directions) Gaussian Markov Random Field (GMRF) is considered wherein $\mathbb{E} [S(\boldsymbol{y},t)] = 0$, $\mathbb{E} [S(\boldsymbol{y},t)^2] = 1$, $\forall \boldsymbol{y},t$. The correlation coefficient between two statuses is assumed to be a function of their Euclidean distance in $\mathbb{R}^d$ and their generation time difference, i.e.,
    \begin{equation}
    \rho(\boldsymbol{y},\boldsymbol{y}',t,t') = \mathbb{E}[S(\boldsymbol{y},t),S(\boldsymbol{y}',t')] \triangleq \operatorname{g}(\|\boldsymbol{y}-\boldsymbol{y}'\|_2,|t-t'|),
    \end{equation}
    where $\operatorname{g}(\|\boldsymbol{y}-\boldsymbol{y}'\|_2,|t-t'|)$ is non-negative, and non-increasing in both components. A GMRF, in addition to being a Gaussian random field, satisfies Markov properties explained as follows.
    \begin{defi}
    	A Gaussian random field that satisfies
    	\begin{equation}
    	\Pr\{\boldsymbol{y}_i|\{{\boldsymbol{y}_j:j \neq i}\}\} = \Pr\{\boldsymbol{y}_i|\{\boldsymbol{y}_j:j \in \mathcal{N}_i\}\}
    	\end{equation}
    	is a GMRF, wherein the points in $\mathcal{N}_i$ are ``close'' to $\boldsymbol{y}_i$.
    \end{defi}
    
    The ``close'' notation in the GMRF definition can be specified based on conditional independence represented by dependency graphs \cite{rue05}. Without going into much details about GMRF which is out of the scope of this paper, we adopt the Nearest Neighbor Graph (NNG), i.e., 
    \begin{equation}
    \mathcal{N}_i \triangleq \{j:\boldsymbol{y}_j=\argmin_{\boldsymbol{y}_j \in \boldsymbol{X}}{\|\boldsymbol{y}_j-\boldsymbol{y}_i\|_2}\}.
    \end{equation}
    Note that under the Poisson assumption, the nearest point is unique with probability $1$, and hence the NNG is well-defined almost surely. Due to Gaussianity, the minimum mean square error estimate of $S(\boldsymbol{y}',t')$ given $S(\boldsymbol{y},t)$ is 
    \begin{equation}
    \mathbb{E}[S(\boldsymbol{y}',t')|S(\boldsymbol{y},t)] = \rho(\boldsymbol{y},\boldsymbol{y}',t,t') S(\boldsymbol{y},t),
    \end{equation}
    with an independent error with power of $1-\rho(\boldsymbol{y},\boldsymbol{y}',t,t')^2$. Combined with the NNG assumption, the average mean squared estimation error in the area is
    \begin{iarray}
    \bar{\xi}_{\mathcal{A}}(t) &\triangleq& \frac{1}{|\mathcal{A}|} \int_{\boldsymbol{y} \in \mathcal{A}} {\xi}({\boldsymbol{y}},t) \mathsf{d}\boldsymbol{y}, \nonumber\\
    {\xi}({\boldsymbol{y}},t) &\triangleq&  \min_{\boldsymbol{x} \in \boldsymbol{X}} \{1-\operatorname{g}^2(\|\boldsymbol{y}-\boldsymbol{x}\|_2,\Delta_{\boldsymbol{x}}(t))\},
    \end{iarray}
    where $\Delta_{\boldsymbol{x}}(t)$ is the AoI at time $t$ for the spatial sampling point $\boldsymbol{x}$, which is defined as $\Delta_{\boldsymbol{x}}(t) = t-u_{\boldsymbol{x}}(t)$ where $u_{\boldsymbol{x}}(t)$ is the generation time of the last successfully received update packet at the fusion center from $\boldsymbol{x}$. 
    
    \textbf{Problem formulation}: The problem of interest is to find the optimal spatial- and time-domain sampling rates that minimize the average remote estimation error of the random field, i.e.,
    \begin{equation}
    	 \{\lambda_\mathsf{s}^*,\lambda_\mathsf{t}^*\} = \argmin_{\lambda_\mathsf{s},\lambda_\mathsf{t}} \left\{\limsup_{T \to \infty} \frac{1}{T} \int_{t=0}^{T} \bar{\xi}_{\mathcal{A}}(t) \mathsf{d}t\right\}.
    \end{equation}
    Hereinafter, we will focus on 1-dimensional space for analytical solutions, i.e., $\mathcal{A} \in \mathbb{R}^{1}$.
    \section{Analytic Results for Exponential Correlation Model and FCFS Discipline}
    \label{sec_mr}
    The analysis in this paper is built upon several recent results in the fields of stochastic geometry and AoI. Before we begin, for concreteness, we select the correlation function $\operatorname{g}(\cdot)$ to be a exponential function of product form, i.e.,
    \begin{equation}
    \label{g_exp}
    \operatorname{g}(\|\boldsymbol{y}-\boldsymbol{y}'\|_2,|t-t'|) = \exp\left(-\frac{b\|\boldsymbol{y}-\boldsymbol{y}'\|_2}{2}  - \frac{a|t-t'|}{2}   \right),
    \end{equation}
    where $a,b\in \mathbb{R}^+$ are constant scaling factors in time and spatial domains, respectively. More involved correlation functions exist such as Mat\'ern covariances \cite{rue05} in GMRF and non-product form functions; however, since the AoI and spatial sampling are independent and isotropic fields are considered such that only the distance matters, it is argued that the presented analysis can be straightforwardly generalized to these correlation functions. In addition, measurements show that real-world random fields such as channel state information obey the time and spatial exponential correlation model \cite{di14}.
    
    Due to the uniformly random spatial sampling---or equivalently Poisson point process---assumption, the spatial statistics of a point in $\mathcal{A}$ has no dependence on its location. Therefore, $\bar{\xi}_{\mathcal{A}}(t) = \frac{1}{|\mathcal{A}|} \int_{\boldsymbol{y} \in \mathcal{A}} {\xi}({\boldsymbol{y}},t) \mathsf{d}\boldsymbol{y} = \mathbb{E} [{\xi}({\boldsymbol{o}},t)]$,	where $\boldsymbol{o}$ denotes an arbitrary point in $\mathcal{A}$. Without loss of generality, we assume $\boldsymbol{o}$ is the origin. In what follows, we derive the CDF of ${\xi}({\boldsymbol{o}},t)$.
	\begin{iarray}
	\label{z_cdf}
	&& \Pr \{{\xi}({\boldsymbol{o}},t) \le z\} \nonumber\\
	&=& 	1 - \Pr \{{\xi}({\boldsymbol{o}},t) \ge z\}\nonumber\\
	&=& 1 - \Pr \{\min_{\boldsymbol{x} \in \boldsymbol{X}} \{1-\operatorname{g}^2(\|\boldsymbol{x}\|_2,\Delta_{\boldsymbol{x}}(t))\} \ge z\} \nonumber\\
	&\overset{(a)}{=}& 1 - \Pr \{ 1-\operatorname{g}^2( \min_{\boldsymbol{x} \in \boldsymbol{X}} \|\boldsymbol{x}\|_2,\Delta_{\boldsymbol{x}_0}(t)) \ge z\} \nonumber\\
	&\overset{(b)}{=}& \Pr\left\{{b} d_\mathsf{min} + a \Delta_{\boldsymbol{x}_0}(t) \le -\log(1-z) \right\},
	\end{iarray}
	where the equality $(a)$ stems from the fact that $\operatorname{g}(\cdot)$ is non-negative and non-increasing and the AoIs of spatial sampling points are i.i.d. since the time-domain sampling is i.i.d. among spatial points and the scheduling of spatial points is uniformly random ($\boldsymbol{x}_0$ represents an arbitrary fixed point). The equality $(b)$ follows from \eqref{g_exp}, and $d_\mathsf{min} \triangleq \min_{\boldsymbol{x} \in \boldsymbol{X}}\|\boldsymbol{x}\|_2$.
	
	Denote $\boldsymbol{D} \triangleq {b} \|d_\mathsf{min}\|_2$, $ \boldsymbol{H} \triangleq a \Delta_{\boldsymbol{x}_0}(t)$ and $\boldsymbol{K} \triangleq \boldsymbol{D}+\boldsymbol{H}$. The CDF of $\boldsymbol{D}$ in $1$-dimensional space and the corresponding Laplace--Stieltjes Transform (LST) are
	\begin{iarray}
		\label{aoi_ur}
	&& F_{\boldsymbol{D}}(d) = 1-e^{-\frac{2\lambda_\mathsf{s}}{b} d}, d \in [0,\infty), \nonumber\\
	&& \mathcal{L}\{F_{\boldsymbol{D}}\}(s) = \frac{2\lambda_\mathsf{s}/b}{s+2\lambda_\mathsf{s}/b},
	\end{iarray}
	respectively. To calculate the CDF of $\boldsymbol{H}$, we have the following lemma. 
	\begin{lemma}
		\label{lm1}
		Considering the $M$ spatial sampling points are scheduled based on a uniformly random manner, the stationary CDF of the AoI of any point and its LST are
		\begin{iarray}
			F^{\mathsf{UR}}_\Delta(t) &=& 1-e^{-(1-\rho_0)\mu_0 t}-\left(\frac{1}{1-\rho_0}+\rho_0 \mu_0 t\right)e^{-\mu_0 t} \nonumber\\
			&& +\frac{1}{1-\rho_0}e^{-\lambda_t t}, \nonumber\\
			\mathcal{L}\{F^{\mathsf{UR}}_\Delta\}(s) &=& \frac{(1-\rho_0)\mu_0}{s+(1-\rho_0)\mu_0} -\frac{(1-\rho_0)\mu_0 s (s+\lambda_t + \mu_0)}{(s+\mu_0)^2(s+\lambda_t)}, \nonumber\\
		\end{iarray}
		wherein $\rho_0 = \lambda_t/\mu_0$. For round-robin scheduling, the LST is
		\begin{iarray}
			&& \mathcal{L}\{F^{\mathsf{RR}}_\Delta\}(s) = w(s) - \frac{(1-\rho_0) s }{s+\lambda_t q(s+\lambda_t)}q(s), \nonumber\\
			&& q(s) = \left(\frac{\mu}{s+\mu}\right)^M,\,w(s)=\frac{(1-\rho_0)s}{s-\lambda_t+\lambda_t q(s)}q(s).
		\end{iarray}
	\end{lemma}
	\begin{IEEEproof}
		A sketch is given in Appendix \ref{app_lm1}.
	\end{IEEEproof}

	Note that uniformly random scheduling is considered hereinafter. Based on \eqref{aoi_ur}, the LST of the CDF of $\boldsymbol{H}$ is $\mathcal{L}\{F_{\boldsymbol{H}}\}(s) = \mathcal{L}\{F^{\mathsf{UR}}_\Delta\}(as)$. Therefore, combining with the fact that the AoI and the spatial point process are independent, the LST of the CDF of $\boldsymbol{K}$ is
    \begin{iarray}
    && \mathcal{L}\{F_{\boldsymbol{K}}\}(s) = \mathcal{L}\{F_{\boldsymbol{D}}\}(s) \cdot \mathcal{L}\{F_{\boldsymbol{H}}\}(s) \nonumber\\
    && =  \frac{\frac{2\lambda_\mathsf{s}}{b}}{s+\frac{2\lambda_\mathsf{s}}{b}} \frac{\frac{(1-\rho_0)\mu_0}{a}}{s+\frac{(1-\rho_0)\mu_0}{a}} -  \frac{\frac{2\lambda_\mathsf{s}}{b}}{s+\frac{2\lambda_\mathsf{s}}{b}} \frac{(1-\rho_0)\mu_0 s (s+\frac{\lambda_t + \mu_0}{a})}{a(s+\frac{\mu_0}{a})^2(s+\frac{\lambda_t}{a})}. \nonumber\\
    \end{iarray}
	The inverse LST of the above formula yields the CDF of $\boldsymbol{K}$: 
	\begin{iarray}
		 F_{\boldsymbol{K}}(x) &=& \mathcal{H}_{\frac{2 \lambda_\mathsf{s}}{b},\frac{(1-\rho_0)\mu_0}{a}}(x) +\frac{1}{1-\rho_0}\mathcal{H}_{\frac{2\lambda_\mathsf{s}}{b},\frac{\lambda_\mathsf{t}}{a}}(x) \nonumber\\
		 && -\frac{1}{1-\rho_0}\mathcal{H}_{\frac{2\lambda_\mathsf{s}}{b},\frac{\mu_0}{a}}(x)+\frac{2\lambda_\mathsf{t}\lambda_\mathsf{s}}{ab} \nonumber\\
		 && \left(-\frac{1}{\left(\frac{2\lambda_\mathsf{s}}{b}-\frac{\mu_0}{a}\right)^2} \mathcal{E}_{\frac{2\lambda_\mathsf{s}}{b}}(x) + \frac{1}{\left(\frac{2\lambda_\mathsf{s}}{b}-\frac{\mu_0}{a}\right)^2} \mathcal{E}_{\frac{\mu_0}{a}}(x) \right.\nonumber\\
		 && \left. + \frac{1}{\frac{2\lambda_\mathsf{s}}{b}-\frac{\mu_0}{a}} xe^{-\frac{\mu_0}{a}x}\right), \\
		 \mathcal{H}_{\lambda_1,\lambda_2}(x) &=& \left\{ \begin{array}{ll}  1-\frac{\lambda_2}{\lambda_2-\lambda_1}e^{-\lambda_1 x}+\frac{\lambda_1}{\lambda_2-\lambda_1}e^{-\lambda_2 x}, &\lambda_1\neq \lambda_2, \\
		 	1-(1+\lambda_1 x)e^{-\lambda_1 x},&\lambda_1=\lambda_2,
		 \end{array} \right.\nonumber\\	 
	 	 \label{H}
	\end{iarray}
	where $\mathcal{E}_{\lambda}(x)$ denote the CDF of an exponential distribution with parameter $\lambda$, i.e., $\mathcal{E}_{\lambda}(x) = 1-e^{-\lambda x}$.
	Combined with \eqref{z_cdf}, we obtain for $z \in [0,1]$, 
	\begin{equation}
	F_{{\xi}({\boldsymbol{o}},t)}(z) = \Pr \{{\xi}({\boldsymbol{o}},t) \le z\} =  F_{\boldsymbol{K}}(-\log(1-z)),
	\end{equation}
	and the Probability Density Function (PDF) follows
	\begin{iarray}
		&& f_{{\xi}({\boldsymbol{o}},t)}(z) \nonumber\\
		&=& \alpha (1-z)^{\frac{2\lambda_\mathsf{s}}{b}-1} + \beta (1-z)^{\frac{\lambda_\mathsf{t}}{a}-1} + \gamma (1-z)^{\frac{\mu_o}{a}-1} \nonumber\\
		&& + \omega \log(1-z) (1-z)^{\frac{\mu_0}{a}-1} + \kappa (1-z)^{\frac{(1-\rho_0)\mu_0}{a}-1}, \nonumber\\		
		\alpha &=& \frac{\frac{(1-\rho_0)\mu_0}{a}\frac{2\lambda_\mathsf{s}}{b}}{\frac{(1-\rho_0)\mu_0}{a}-\frac{2\lambda_\mathsf{s}}{b}} + \frac{1}{1-\rho_0}\left(\frac{\frac{\lambda_\mathsf{t}}{a}\frac{2\lambda_\mathsf{s}}{b}}{\frac{\lambda_\mathsf{t}}{a}-\frac{2\lambda_\mathsf{s}}{b}} - \frac{\frac{\mu_0}{a}\frac{2\lambda_\mathsf{s}}{b}}{\frac{\mu_0}{a}-\frac{2\lambda_\mathsf{s}}{b}} \right) \nonumber\\
		&& - \frac{4\lambda_\mathsf{t} \lambda_\mathsf{s}^2}{ab^2\left(\frac{\mu_0}{a}-\frac{2\lambda_\mathsf{s}}{b}\right)^2},\,\kappa=-\frac{\frac{2\lambda_\mathsf{s}}{b}\frac{(1-\rho_0)\mu_0}{a}}{\frac{(1-\rho_0)\mu_0}{a}-\frac{2\lambda_\mathsf{s}}{b}}, \nonumber\\
		\beta &=& - \frac{1}{1-\rho_0}\frac{\frac{2\lambda_\mathsf{s}}{b}\frac{\lambda_\mathsf{t}}{a}}{\frac{\lambda_\mathsf{t}}{a}-\frac{2\lambda_\mathsf{s}}{b}},\, \omega = -\frac{2\lambda_\mathsf{t} \lambda_\mathsf{s}\frac{\mu_0}{a}}{ab\left(\frac{\mu_0}{a}-\frac{2\lambda_\mathsf{s}}{b}\right)},\nonumber\\
		\gamma &=& \frac{1}{1-\rho_0}\frac{\frac{2\lambda_\mathsf{s}}{b}\frac{\mu_0}{a}}{\frac{\mu_0}{a}-\frac{2\lambda_\mathsf{s}}{b}} +  \frac{4\lambda_\mathsf{t} \lambda_\mathsf{s}^2}{ab^2\left(\frac{\mu_0}{a}-\frac{2\lambda_\mathsf{s}}{b}\right)^2},
	\end{iarray}
	when the set $\left\{\frac{2\lambda_\mathsf{s}}{b}, \frac{\lambda_\mathsf{t}}{a}, \frac{\mu_o}{a}, \frac{(1-\rho_0)\mu_0}{a}\right\}$ has no repetitive entries; otherwise the PDF can be obtained based on the second case in \eqref{H} with simple mathematical manipulations. According to definite integrals such as $\int_{0}^{1}x^{p-1}(1-x)^{q-1} \mathsf{d}x = \frac{\Gamma(p)\Gamma(q)}{\Gamma(p+q)}$ for $p,q>0$ where $\Gamma(\cdot)$ is the Gamma function, and $\int_{0}^{\infty}x^{n}e^{-px} \mathsf{d}x = \frac{n!}{p^{n+1}}$ for $p>0,n=1,2,\cdots$, we can obtain the average remote estimation error of the random field 
	\begin{iarray}
		\label{e_fcfs}
		&& \bar{\epsilon} \triangleq \mathbb{E}\left\{\limsup_{T \to \infty} \frac{1}{T} \int_{t=0}^{T} \bar{\xi}_{\mathcal{A}}(t) \mathsf{d}t \right\} \nonumber\\
		&=& \frac{\alpha}{\frac{2\lambda_\mathsf{s}}{b}{\left(\frac{2\lambda_\mathsf{s}}{b}+1\right)}} + \frac{\beta}{\frac{\lambda_\mathsf{t}}{a}{\left(\frac{\lambda_\mathsf{t}}{a}+1\right)}} + \frac{\gamma}{\frac{\mu_0}{a}{\left(\frac{\mu_0}{a}+1\right)}} \nonumber\\
		&& + \omega \left(\frac{1}{\left(\frac{\mu_0}{a}+1\right)^2}-\frac{1}{\left(\frac{\mu_0}{a}\right)^2}\right) + \frac{\kappa}{\frac{(1-\rho_0)\mu_0}{a}{\left(\frac{(1-\rho_0)\mu_0}{a}+1\right)}} \nonumber\\
		&=& 1-\mathcal{F}\left(\frac{2\lambda_\mathsf{s}}{b}\right) \mathcal{F}\left(\frac{\lambda_\mathsf{t}}{a}\right) \mathcal{F} \left(\frac{\bar{\mu}}{\lambda_\mathsf{s}a}-\frac{\lambda_\mathsf{t}}{a}\right) \nonumber\\
		&& \cdot\left[1+\frac{\frac{\lambda_\mathsf{t}}{a}}{\left(\frac{\bar{\mu}}{\lambda_\mathsf{s}a}+1\right)^2}\right], 
	\end{iarray}
	where $\mathcal{F}\left(x\right) \triangleq \frac{x}{x+1}$, and $\frac{\bar{\mu}}{\lambda_\mathsf{s}a}-\frac{\lambda_\mathsf{t}}{a}>0$. Observing this formula, when $\frac{\lambda_\mathsf{t}}{a} \to 0$, $\bar{\epsilon} \to 1$; similarly, $\bar{\epsilon} \to 1$ when $\frac{\bar{\mu}}{\lambda_\mathsf{s}a}-\frac{\lambda_\mathsf{t}}{a} \to 0$, $\frac{\bar{\mu}}{\lambda_\mathsf{s}a}-\frac{\lambda_\mathsf{t}}{a} \to \infty$ or $\frac{\lambda_\mathsf{t}}{a}  \to \infty$. Therefore, the global minimum of $\bar{\epsilon}$ happens at a point where $\frac{\partial{\bar{\epsilon}}}{\partial{\lambda_\mathsf{s}}}=0$ and $\frac{\partial{\bar{\epsilon}}}{\partial{\lambda_\mathsf{t}}}=0$. 
	
\begin{figure}[!t]
		\centering    
		{\includegraphics[width=0.5\textwidth]{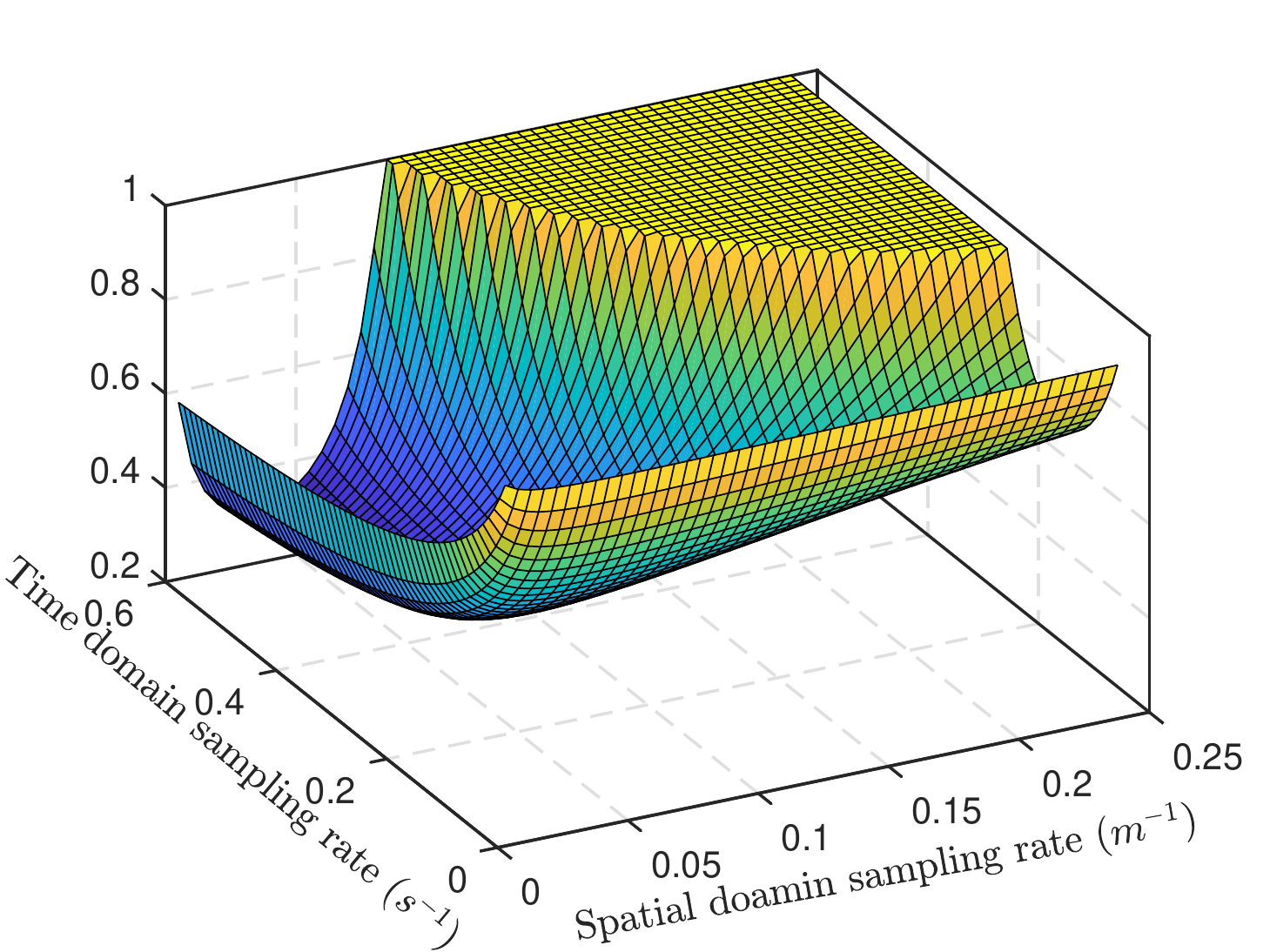}}
		\caption{Visualization of the error function \eqref{e_fcfs}.}
		\label{fig_2d}
	\end{figure}

	Based on a numerical result, Fig. \ref{fig_2d} shows the remote estimation error variation with the spatial and time sampling rates $\lambda_\mathsf{s}$ and $\lambda_\mathsf{t}$. The area is a one-dimensional line with length of $1000$~m. The parameters in the exponential correlation function are $a=10^{-1}$ and $b=10^{-2}$. The service rate is $\mu = 0.05$~$s^{-1}$. The time- and spatial sampling rates are $\lambda_\mathsf{t}=0.2$~$s^{-1}$ and $\lambda_\mathsf{s}=0.1$~$m^{-1}$ in the corresponding subfigure. It is observed that the simulation and analytical results coincide exactly, and that there exists an optimal point $(\lambda_\mathsf{t}^*,\lambda_\mathsf{s})^*$ where the error is minimized; although an explicit expression seems elusive, the exact location of the optimal point can be found by a two-dimensional search method. A visualization of the error function is presented in Fig. \ref{fig_2d}.
	\begin{figure}[!t]
		\centering    
		\subfigure{\includegraphics[width=0.235\textwidth]{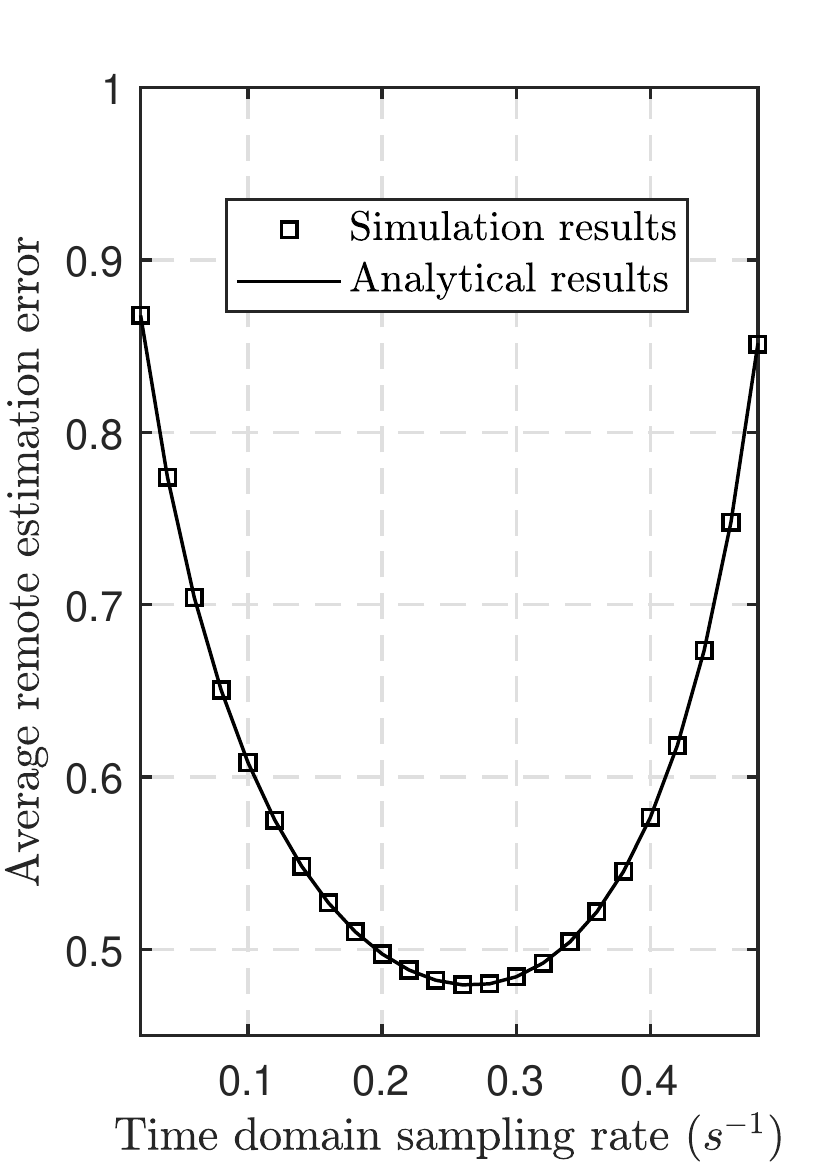}}
		\subfigure{\includegraphics[width=0.235\textwidth]{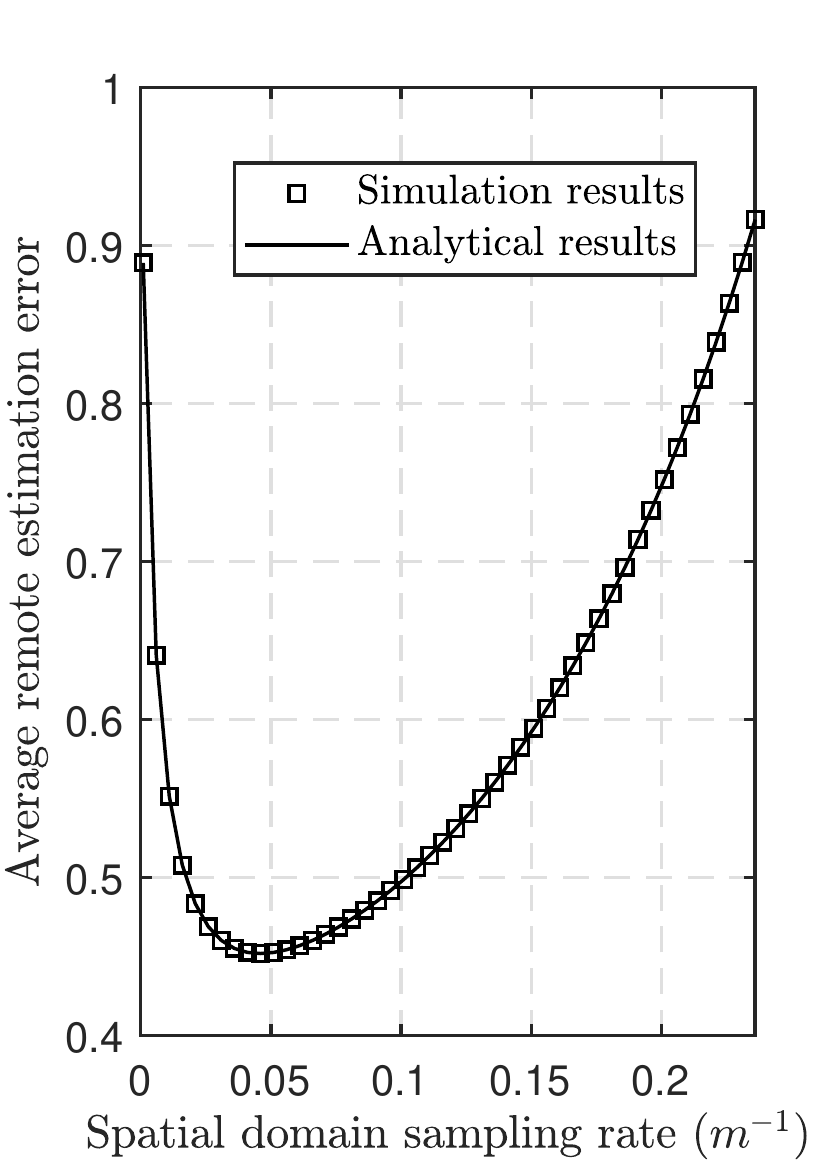}}
		\caption{Simulation and analytical results for remote estimation error. }
		\label{fig_sim}
	\end{figure} 

	\section{Analytic Results for Exponential Correlation Model and LCFS Discipline}
    \label{sec_pm}
    In the scenario wherein each spatial sampling point updates their packets based on a LCFS discipline and no preemption, which is obviously the optimal packet management policy at each point, the AoI stationary distribution can be obtained based on the following lemma. Note that queue stability is not considered here, i.e., the queuing system at each sampling point is equivalent to only keeping the freshest packet to date. 
    \begin{lemma}
    	\label{lm2}
		Considering uniformly random scheduling among points, the AoI stationary distribution of every spatial sampling point with LCFS discipline and the corresponding LST are
		\begin{iarray}
			\hat{F}^{\mathsf{UR}}_\Delta(t) &=& 1-\frac{1}{1-\rho_0}e^{- {\frac{\lambda_\mathsf{t}}{a}} x} + \frac{1}{1-\rho_0}e^{-{\mu_0} x}, \nonumber\\
			\mathcal{L}\{\hat{F}^{\mathsf{UR}}_\Delta\}(s) &=&  \frac{{\lambda_\mathsf{t}}}{s+{\lambda_\mathsf{t}}} \frac{{\mu_0}}{s+{\mu_0}}. 
		\end{iarray}	
    \end{lemma}
    \begin{IEEEproof}
    The proof is based on an arrival theorem in queuing systems \cite{jiang18_iot}. Details are omitted.
    \end{IEEEproof}
    
    Similarly with \eqref{z_cdf}, define $\hat{\boldsymbol{K}}$ = $\hat{\boldsymbol{D}}+\hat{\boldsymbol{H}}$ where $\hat{\boldsymbol{D}} = \boldsymbol{D}$ and $\hat{\boldsymbol{H}}$ can be given by Lemma \ref{lm2}. It follows that, $\mathcal{L}\{F_{\hat{\boldsymbol{K}}}\}(s) = \frac{\frac{2\lambda_\mathsf{s}}{b}}{s+\frac{2\lambda_\mathsf{s}}{b}} \frac{\frac{\lambda_\mathsf{t}}{a}}{s+\frac{\lambda_\mathsf{t}}{a}} \frac{\frac{\mu_0}{a}}{s+\frac{\mu_0}{a}}$,	and the average remote estimation error is
	\begin{iarray}
		\label{err_one}
		\bar{\epsilon} &=& \frac{\eta}{\frac{2\lambda_\mathsf{s}}{b}+1} + \frac{\nu}{\frac{\lambda_\mathsf{t}}{a}+1} + \frac{\upsilon}{\frac{\mu_0}{a}+1},  \nonumber\\
		\eta &=& \frac{\frac{\lambda_\mathsf{t}}{a} \frac{\mu_0}{a}}{\left(\frac{2\lambda_\mathsf{s}}{b}-\frac{\lambda_\mathsf{t}}{a}\right)\left(\frac{2\lambda_\mathsf{s}}{b}-\frac{\mu_0}{a}\right)}, \nonumber\\
		\nu &=& \frac{\frac{2\lambda_\mathsf{s}}{b} \frac{\mu_0}{a}}{\left(\frac{\lambda_\mathsf{t}}{a}-\frac{2\lambda_\mathsf{s}}{b}\right)\left(\frac{\lambda_\mathsf{t}}{a}-\frac{\mu_0}{a}\right)},\, \upsilon = \frac{\frac{2\lambda_\mathsf{s}}{b} \frac{\lambda_\mathsf{t}}{a}}{\left(\frac{\mu_0}{a}-\frac{2\lambda_\mathsf{s}}{b}\right)\left(\frac{\mu_0}{a}-\frac{\lambda_\mathsf{t}}{a}\right)}. 
	\end{iarray}
	
	To obtain the optimal spatial and time sampling rates, we first note that, in this case of only keeping the freshest packet, the optimal time domain sampling rate should be as large as possible, i.e., $\lambda_\mathsf{t}^* \to \infty$. The optimal spatial sampling rate can be derived as follows. When $\lambda_\mathsf{t}^* \to \infty$ in \eqref{err_one},
	\begin{equation}
		\lim\limits_{\lambda_\mathsf{t}\to \infty}\bar{\epsilon} = 1-\mathcal{F}\left(\frac{2\lambda_\mathsf{s}}{b}\right) \mathcal{F}\left(\frac{\mu_0}{a}\right) \ge 1-\mathcal{F}^2\left(\sqrt{\frac{2\bar{\mu}}{ab}}\right).
	\end{equation}
	The optimal spatial sampling rate is therefore $\lambda_\mathsf{s}^* = \sqrt{\frac{b \bar{\mu}}{2a}}$.
		
    \section{Conclusions and Future Work}
    \label{sec_cl}
    In this paper, the remote estimation of status information with spatial structures is investigated. The status information is described by a GMRF in the spatial domain. Closed-form expressions for remote estimation error with FCFS and LCFS disciplines in the one-dimensional GMRF were both obtained, assuming an exponential correlation function. It is found that the derived optimal time- and spatial-sampling rates can optimize the remote estimation accuracy, by balancing the tradeoff between information distance (in both time and spatial domains) and network queuing delay. 
    
    Future work includes generalizations to 2-dimensional (2D) and 3-dimensional (3D) random fields, and non-Gaussian, non-Markov random fields. We note that the difficulty of generalizations to 2D and 3D random fields is mainly mathematical, i.e., the Laplace transform of the CDF becomes very complicated in 2D and 3D cases, while the methodology is still effective. However, the extension to non-Gaussian and non-Markov random fields may need novel methodologies.
%    \section*{Acknowledgments}
    
    \appendices
    \section{Proof of Lemma \ref{lm1}}
    \label{app_lm1}
    The proof is mainly based on the results in \cite{ino18}. Note that the service time for each $\boldsymbol{x}$ with uniformly random scheduling is exponentially distributed with mean rate of $\mu_0$, which can be intuitively explained as follows: since the service time is still memoryless given uniformly random scheduling and exponentially distributed transmission time, and the fact that the only memoryless continuous distribution on $(0,\infty)$ is exponential distribution, the service time for each point should obey exponential distribution. Formally, it can be deduced based on a summation of conditional probability with Erlang distributed random variables, which is omitted due to lack of space. The situation for round-robin scheduling is similar, expect that the service time obeys Erlang distribution. Then, we invoke the results in \cite[Theorem 25]{ino18} to obtain the LST and stationary CDF. The details are again omitted.
    \bibliographystyle{ieeetr}
    \bibliography{spatial_aoi}

\begin{thebibliography}{10}

\bibitem{villa16}
T.~F. Villa, F.~Salimi, K.~Morton, L.~Morawska, and F.~Gonzalez, ``Development
  and validation of a {UAV} based system for air pollution measurements,'' {\em
  Sensors}, vol.~16, no.~12, p.~2202, 2016.

\bibitem{oll07}
A.~Ollero, P.~J. Marron, M.~Bernard, J.~Lepley, M.~la~Civita, E.~de~Andres, and
  L.~van Hoesel, ``{AWARE}: Platform for autonomous self-deploying and
  operation of wireless sensor-actuator networks cooperating with unmanned
  aerial vehicles,'' in {\em 2007 IEEE International Workshop on Safety,
  Security and Rescue Robotics}, pp.~1--6, Sep 2007.

\bibitem{deng18}
R.~Deng, Z.~Jiang, S.~Zhou, S.~Cui, and Z.~Niu, ``A two-step learning and
  interpolation method for location-based channel database,'' in {\em IEEE
  Global Commun. Conf. (GLOBECOM)}, Dec 2018.

\bibitem{kaul12}
S.~Kaul, R.~Yates, and M.~Gruteser, ``Real-time status: How often should one
  update?,'' in {\em IEEE INFOCOM}, pp.~2731--2735, Mar 2012.

\bibitem{costa16}
M.~Costa, M.~Codreanu, and A.~Ephremides, ``On the age of information in status
  update systems with packet management,'' {\em IEEE Trans. Inform. Theory},
  vol.~62, pp.~1897--1910, April 2016.

\bibitem{najm17}
E.~Najm, R.~Yates, and E.~Soljanin, ``Status updates through {M/G/1/1} queues
  with {HARQ},'' in {\em IEEE Int'l Symp. Info. Theory}, pp.~131--135, Jun
  2017.

\bibitem{sun17}
Y.~Sun, E.~Uysal-Biyikoglu, R.~Yates, C.~E. Koksal, and N.~B. Shroff, ``Update
  or wait: How to keep your data fresh,'' in {\em IEEE INFOCOM}, pp.~1--9,
  April 2016.

\bibitem{yin18}
Y.~Sun and B.~Cyr, ``Information aging through queues: A mutual information
  perspective,'' in {\em IEEE SPAWC Conference}, 2018.

\bibitem{kadota18}
I.~Kadota, A.~Sinha, E.~Uysal-Biyikoglu, R.~Singh, and E.~Modiano, ``Scheduling
  policies for minimizing age of information in broadcast wireless networks,''
  {\em arXiv preprint arXiv:1801.01803}, 2018.

\bibitem{hsu18}
Y.-P. Hsu, ``Age of information: {Whittle} index for scheduling stochastic
  arrivals,'' in {\em IEEE Int'l Symp. Info. Theory}, 2018.

\bibitem{jiang18_iot}
Z.~Jiang, B.~Krishnamachari, X.~Zheng, S.~Zhou, and Z.~Niu, ``Timely status
  update in wireless uplinks: Analytical solutions with asymptotic
  optimality,'' {\em IEEE Internet of Things Journal}, 2018.

\bibitem{jiang18_itc}
Z.~Jiang, B.~Krishnamachari, S.~Zhou, and Z.~Niu, ``Can decentralized status
  update achieve universally near-optimal age-of-information in wireless
  multiaccess channels?,'' in {\em International Teletraffic Congress (ITC
  30)}, Sep 2018.

\bibitem{rajat19}
V.~Tripathi, R.~Talak, and E.~Modiano, ``Age optimal information gathering and
  dissemination on graphs,'' in {\em IEEE INFOCOM}, pp.~1--9, April 2019.

\bibitem{jiang19}
Z.~Jiang, S.~Zhou, Z.~Niu, and Y.~Cheng, ``A unified sampling and scheduling
  approach for status update in wireless multiaccess networks,'' in {\em IEEE
  INFOCOM}, pp.~1--9, April 2019.

\bibitem{hri18}
J.~Hribar, M.~Costa, N.~Kaminski, and L.~A. DaSilva, ``Using correlated
  information to extend device lifetime,'' {\em IEEE Internet of Things
  Journal}, pp.~1--1, 2018.

\bibitem{baccelli10}
F.~Baccelli, B.~B{\l}aszczyszyn, {\em et~al.}, ``Stochastic geometry and
  wireless networks: {Volume I Theory},'' {\em Foundations and
  Trends{\textregistered} in Networking}, vol.~3, no.~3--4, pp.~249--449, 2010.

\bibitem{rue05}
H.~Rue and L.~Held, {\em {Gaussian Markov} random fields: theory and
  applications}.
\newblock CRC press, 2005.

\bibitem{di14}
R.~Di~Taranto, S.~Muppirisetty, R.~Raulefs, D.~Slock, T.~Svensson, and
  H.~Wymeersch, ``Location-aware communications for {5G} networks: How location
  information can improve scalability, latency, and robustness of {5G},'' {\em
  IEEE Signal Process. Mag.}, vol.~31, no.~6, pp.~102 -- 112, 2014.

\bibitem{ino18}
Y.~Inoue, H.~Masuyama, T.~Takine, and T.~Tanaka, ``A general formula for the
  stationary distribution of the age of information and its application to
  single-server queues,'' {\em arXiv preprint arXiv:1804.06139}, 2018.

\end{thebibliography}
    \end{document}